\documentclass[showpacs,amsmath,
preprintnumbers,
preprint,endfloats*,
aps,prl]{revtex4}
\usepackage{graphicx}
\usepackage{dcolumn}
\usepackage[colorlinks=true, pdfstartview=FitV, linkcolor=blue,
citecolor=blue, urlcolor=blue]{hyperref}

\def \ie {{\it i.e.\ }}
\def \etal {{\it et al.}}

\newcommand{\NI}{N_I}
\newcommand{\NS}{N_S}
\newcommand{\DNI}{n_I}
\newcommand{\DNS}{n_S}
\newcommand{\SDNI}{S_{\DNI}}

\newcommand{\kI}{k_I}
\newcommand{\kS}{k_S}
\newcommand{\kIS}{k_{IS}}
\newcommand{\rI}{r_I}
\newcommand{\rS}{r_S}
\newcommand{\rIS}{r_{IS}}
\newcommand{\SrI}{S_{\rI}}
\newcommand{\SrS}{S_{\rS}}
\newcommand{\SrIS}{S_{\rIS}}

\newcommand{\BI}{B_{1I}}
\newcommand{\BS}{B_{1S}}
\newcommand{\BeffI}{{\bf B}_{\rm{eff},I}}
\newcommand{\BeffS}{{\bf B}_{\rm{eff},S}}
\newcommand{\BeffH}{{\bf B}_{\rm{eff},H}}
\newcommand{\BeffC}{{\bf B}_{\rm{eff},C}}
\newcommand{\tmI}{\tau_{I}}
\newcommand{\tmS}{\tau_{S}}
\newcommand{\tcp}{\tau_{IS}}

\begin{document}

\title{Force-detected nuclear double resonance between statistical
  spin polarizations} \author{M. Poggio$^{1,2}$, H. J. Mamin$^1$, C.
  L.  Degen$^1$, M.  H. Sherwood$^1$, and D. Rugar$^1$}
\affiliation{$^1$IBM Research Division,
  Almaden Research Center, 650 Harry Rd., San Jose CA\\
  $^2$Center for Probing the Nanoscale, Stanford University, 476
  Lomita Hall, Stanford CA} \date{\today}

\begin{abstract}
  We demonstrate nuclear double resonance for nanometer-scale volumes
  of spins where random fluctuations rather than Boltzmann
  polarization dominate.  When the Hartmann-Hahn condition is met in a
  cross-polarization experiment, flip-flops occur between two species
  of spins and their fluctuations become coupled.  We use magnetic
  resonance force microscopy to measure this effect between $^1$H and
  $^{13}$C spins in $^{13}$C-enriched stearic acid.  The development
  of a cross-polarization technique for statistical ensembles adds an
  important tool for generating chemical contrast in nanometer-scale
  magnetic resonance.

\end{abstract}


\maketitle

The physics of microscopic spin ensembles can be distinctly different
from that of macroscopic ensembles.  For example, in volumes of
nuclear spins smaller than about (100~nm)$^3$, random spin flips
generate a fluctuating polarization that exceeds the typical thermal
(or Boltzmann) polarization \cite{Bloch:1946,Sleator:1985,Mamin:2005}.
These spin fluctuations are a major source of dephasing in solid-state
quantum systems \cite{Childress:2006,Reilly:2008}, and their control
is an important prerequisite for nanometer-scale magnetic resonance
imaging (MRI) and spectroscopy
\cite{Mamin:2007,Degen:2007,DegenTMV:2008,DegenAPL:2008,
  Maze:2008,Balasubramanian:2008}.  Recent experiments using magnetic
resonance force microscopy (MRFM) \cite{Rugar:2004,Kuehn:2008} have
extracted useful information from random polarization and harnessed it
for nanometer-scale three-dimensional imaging \cite{DegenTMV:2008}.
One way to further improve nanometer-scale MRI is to combine its
imaging capability with the chemical selectivity intrinsic to magnetic
resonance.  Techniques such as nuclear magnetic resonance (NMR)
spectroscopy routinely exploit this feature for the structural
analysis of molecules in millimeter-sized samples.  At the
nanometer-scale, the same information could be used to locally probe
the chemical composition of materials, and to identify specific
molecules in complex biological structures.

Here we apply nuclear double resonance to achieve such a form of
contrast using cross-polarization (CP) between statistically polarized
$^1$H and $^{13}$C spins in $^{13}$C-enriched stearic acid.  CP is
widely used in NMR for the signal enhancement of low-abundance and
low-$\gamma$ nuclei and forms the basis for many advanced
multidimensional spectroscopy techniques \cite{SlichterCP:1990}.
Indeed, CP has been demonstrated as an efficient chemical contrast
mechanism for micrometer-scale one-dimensional MRFM imaging based on
Boltzmann polarization \cite{Lin:2006,Eberhardt:2007,Eberhardt:2008}.
This spectroscopic method is not directly applicable to statistically
polarized volumes of spins since at any given time the polarization
has a random sign and magnitude, making measured signals intrinsically
irreproducible.  One way around this problem, as demonstrated here, is
to observe the change in the correlation time of the fluctuations
\cite{Mamin:2005,PoggioAPL:2007}.

CP relies on matching the rotating-frame Zeeman splittings of two
different spin species (denoted as $I$ and $S$) in order to promote
cross-species spin flip-flops through the heteronuclear dipolar
coupling \cite{SlichterCP:1990}.  For this purpose, two strong rf
fields are applied with frequencies near the Larmor resonance of the
respective spins.  According to the original work by Hartmann and
Hahn, efficient transfer is achieved when the rf field strengths are
such that the respective Rabi frequencies have a similar magnitude
\cite{Hartmann:1962, SlichterCP:1990}.  The Rabi frequency is
determined by the effective field in the rotating frame; for a spin
$I$, the effective field is given by $\BeffI = \BI \hat{x}' +
\frac{2\pi}{\gamma_{I}} \Delta \nu_{I} \hat{z}$, where $\BI$ is the
magnitude and $\Delta \nu_{I} = \nu_I - \frac{\gamma_I}{2 \pi} B_0$ is
the resonance offset of the rf field with a frequency $\nu_I$.
$\gamma_{I}$ is the nuclear gyromagnetic ratio of $I$, $\hat{x}'$ is a
unit vector in the rotating frame, and $\hat{z}$ is a unit vector
along the static field ${\bf B}_0$ (likewise for the $S$ spin).  The
Hartmann-Hahn (HH) condition can then be expressed as $\gamma_{I}
\left| \BeffI \right| = \gamma_{S} \left | \BeffS \right|$.  The
efficiency of CP depends on the angles of the effective fields with
respect to the $\hat{x}'$ axis in the respective rotating frames, so
that the most efficient transfer occurs when $\Delta \nu_I = \Delta
\nu_S = 0$ and $\gamma_H B_{1H} = \gamma_C B_{1C}$ \cite{Lin:2006}.
For samples with a high spin density and strong nuclear moments, the
transfer process is typically very efficient, occurring on a
characteristic time-scale set by the dipolar coupling frequency
between spins.  In stearic acid it is estimated at about 16 $\mu$s
\cite{NoteTauCP,Xiaoling:1988}.

For Boltzmann polarizations, a spin temperature description is
commonly used in which the two spin polarizations are viewed as
thermal ensembles \cite{SlichterSpinTemp:1990}.  This model
successfully explains the dynamics of CP at the macroscopic level as
well as in micrometer-scale MRFM \cite{Eberhardt:2007}.  When the HH
condition is met, the two spin ensembles come into thermal contact and
their temperatures equilibrate.  In this way, one thermal ensemble can
be used to enhance or deplete the polarization of the other ensemble.

For statistically polarized nuclear spins, the same exchange of
polarization occurs, except that the mean polarization of both spin
ensembles is zero.  When there is no HH contact, the polarization of
an ensemble in the rotating frame fluctuates about zero with a
variance that is given by the number of spins in that ensemble. The
fluctuations in each ensemble occur on a time-scale determined by that
ensemble's rotating frame relaxation rate.  When the HH condition is
met, double resonance allows rapid flip-flop processes that exchange
polarization between the $I$ and $S$ spin ensembles, while
simultaneously conserving the total polarization.  As a result,
fluctuations can occur on a much faster time-scale.

To obtain a more quantitative picture, we consider the case for a
spin-1/2 system where the polarizations of the ensembles are
represented by $\DNI$ and $\DNS$, defined as the difference between
spin-up and spin-down populations for the $I$ and $S$ spins
respectively.  The dynamics of the HH transfer can then be described
by a set of detailed balance equations that use spin population
difference instead of spin temperature (see Ref.
\cite{Hartmann:1962}, Eq. (8)),
\begin{eqnarray}
\frac{\partial\DNI}{\partial t} & = & -\kI\DNI - \frac{\kIS}{N} \left
(\NS\DNI-\NI\DNS\right ) + \rI(t) + \rIS(t), \label{eq:system1}\\
\frac{\partial\DNS}{\partial t} & = & -\kS\DNS + \frac{\kIS}{N} \left
(\NS\DNI-\NI\DNS\right ) + \rS(t) - \rIS(t),
\label{eq:system2}
\end{eqnarray}
where $N$ is the total number of spins in the ensemble, $\NI$ and
$\NS$ are the number of $I$ and $S$ spins respectively ($N=\NI+\NS$),
$\kI$ and $\kS$ are the spin-lattice relaxation rates, and $\kIS$ is
the average rate of exchange between $I$ and $S$ spins.

The original work by Hartmann and Hahn (Ref. \cite{Hartmann:1962}, Eq.
(8)) does not account for statistical spin fluctuations.  To describe
the random excitations that lead to statistical polarization, we
therefore introduce three stochastic functions $\rI(t)$, $\rS(t)$ and
$\rIS(t)$, equivalent to the number of $I$-spin flips, $S$-spin flips,
and cross-species spin flip-flops per unit time, respectively.  We
assume white spectral densities for these functions and further
require that on sufficiently long time-scales, the variance of the
fluctuating polarizations is equal to the number of spins in the
ensemble \cite{Bloch:1946,Degen:2007}.  The resulting (double-sided)
spectral densities are $\SrI = 2 \kI \NI$ and $\SrS = 2 \kS \NS$.  The
spectral density $\SrIS = 2 \kIS \NI \NS / N$ is similarly obtained
\cite{footnote}.

The coupled differential equations (\ref{eq:system1}) and
(\ref{eq:system2}) can be solved in frequency space, yielding
expressions for the spectral densities of $\DNI$ and $\DNS$.  One can
then calculate experimental parameters, such as the variance or the
characteristic time-scale of the spin fluctuations in the presence of
CP\@.  We concentrate our analysis on the spectral density of the
$I$-spin fluctuations:
\begin{equation}
\SDNI =  \frac{ 2 \NI \left [ \left (\kI +  \kIS \frac{\NS}{N}
    \right )\omega^2 + \left (\kS + \kIS \frac{\NI}{N} \right )\left
    [\kI \kS + \kIS \left (\kI \frac{\NI}{N} + \kS \frac{\NS}{N} \right ) \right] \right ]
    }{\omega^4 + \left [\kI^2 + \kS^2 + \kIS^2 + 2 \kIS \left (\kI
    \frac{\NS}{N} + \kS \frac{\NI}{N} \right ) \right
    ]\omega^2 + \left [\kI \kS + \kIS \left
    (\kI \frac{\NI}{N} + \kS \frac{\NS}{N}\right ) \right ]^2 }.
\label{eq:specdens}
\end{equation}
Evaluation of (\ref{eq:specdens}) allows us to determine the behavior
of the $I$-spin fluctuations for arbitrary values of $\kI$, $\kS$ and
$\kIS$.  We consider two particularly relevant cases.  As expected, in
the regime of negligible CP where $\kIS \ll \kI$, $\kS$, the $I$-spin
fluctuations occur on a time-scale $\tmI = \kI^{-1}$ with a variance
equal to $\NI$: $\displaystyle \lim_{\stackrel{\scriptscriptstyle \kIS/\kI \to
    0}{\scriptscriptstyle \kIS/\kS \to 0}} \SDNI = \frac{2 \tmI \NI}{1+ \omega^2
  \tmI^2}$.

In the regime of strong CP where $\kIS \gg \kI$, $\kS$, two
time-scales emerge.  Rapid polarization transfer between spin
ensembles leads to fast spin fluctuations with a characteristic time
$\tcp = \kIS^{-1}$ and a variance $\NI \NS / N$.  The reduction of the
variance compared to $\NI$ occurs because the total polarization is
conserved on time-scales short compared to $\tmI$ and $\tmS =
\kS^{-1}$, thereby limiting the phase space of possible states.  On
top of the rapid exchange of polarization between spin ensembles, the
polarization also fluctuates on a much slower time-scale $\tau_{avg} =
\left (\kI \frac{\NI}{N} + \kS \frac{\NS}{N} \right )^{-1}$ with a
variance given by $\NI^2 / N$.  Note that the sum of the rapid and
slow variances is once again $\NI$.  It is not surprising, then, that
in this strong CP limit the $I$-spin spectral density approaches the
sum of two spectral densities, each with one of the two characteristic
times and variances:
\begin{equation}
\lim_{\stackrel{\scriptscriptstyle \kI/\kIS \to 0}{\scriptscriptstyle \kS/\kIS \to 0}} \left [\SDNI \right ] =
\lim_{\stackrel{\scriptscriptstyle \kI/\kIS \to 0}{\scriptscriptstyle \kS/\kIS \to 0}} \left [\frac{2 \tcp \NI \NS / N}{1 +
    \omega^2 \tcp^2} + \frac{2 \tau_{avg} \NI^2 / N}{1 + \omega^2
    \tau_{avg}^2}\right ].
\label{eq:2ndLimit}
\end{equation}
Thus, the main signature of CP is the presence of spin fluctuations
faster than $\tmI$.  Simple coin-flipping simulations support these
findings.

\begin{figure}[b]\includegraphics{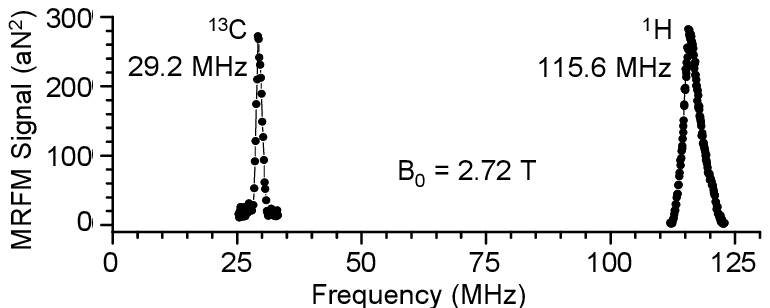}
\caption{\label{fig1}
  MRFM signal as a function of rf frequency at $B_0 = 2.72$ T for both
  the $^1$H and $^{13}$C isotopes in 99\% $^{13}$C-enriched stearic
  acid.  The two data sets were taken under slightly different
  experimental conditions, making their relative amplitudes arbitrary.
}\end{figure}

We demonstrate nuclear CP between statistically polarized $^1$H and
$^{13}$C spins in an experiment using a custom-built magnetic
resonance force microscope \cite{PoggioAPL:2007}.  For these species,
where $\gamma_H / \gamma_C = 3.9772$, the combination of the minimum
$\BI$ needed for adiabatic inversions and the maximum allowed current
in our microfabricated rf field source prevent us from reaching the
condition of most efficient CP, $\gamma_H B_{1H} = \gamma_C B_{1C}$.
Nevertheless, we are able to observe significant CP at non-zero
resonance offsets where the HH condition is met.

The sample is a 10-$\mu$m-sized particle of stearic acid,
C$_{18}$H$_{36}$O$_{2}$, where $>$99\% of the carbon is $^{13}$C.  The
particle is placed upon an ultrasensitive, single-crystal Si
cantilever (120 $\mu$m long, 3 $\mu$m wide, and 0.1 $\mu$m thick),
with part of the particle sticking out beyond the end of the
cantilever.  The cantilever and particle are then briefly heated to a
few $^{\circ}$C below the melting point promoting adhesion to the
cantilever.  The cantilever with the sample attached is then mounted
in a vacuum chamber (pressure $<1 \times 10^{-6}$ torr) at the bottom
of a cryostat, which is isolated from environmental vibrations.  At
the operating temperature of 4.2 K, the sample-loaded cantilever has a
resonant frequency $f_0 = 1 / T_0 = 2.9$ kHz, an intrinsic quality
factor $Q_0 = 44000$, and a spring constant $k = 86$ $\mu$N/m.  We
actively damp the cantilever in order to give it a fast response time
of $\sim 25$ ms.  An FeCo nanomagnetic tip is used to produce the
large ($\sim 10^6$ T/m) spatial field gradient required for generating
magnetic forces of a few attonewtons between the spins in the sample
and the tip.  A microwire underneath the tip generates an rf field of
a few millitesla that induces magnetic resonance in the sample
\cite{PoggioAPL:2007}.  The MRFM measurement is carried out on a
stearic acid sample positioned 100 nm above the magnetic tip in an
externally applied magnetic field $|{\bf B}_{ext}|$ = 2.64 T.

\begin{figure}\includegraphics{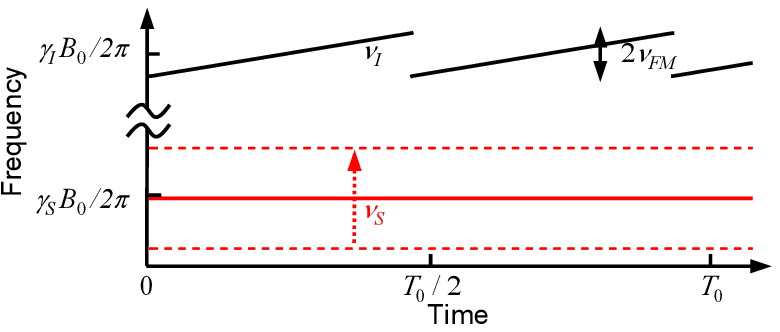}
\caption{\label{fig2}
  A schematic diagram showing the rf frequencies $\nu_I$ and $\nu_S$
  during an experiment.  With $\nu_S$ constant, periodic sweeps of
  $\nu_I$ through resonance adiabatically invert the $I$ spins.  The
  effect of incrementing $\nu_S$ is represented by the dotted arrow.
}\end{figure}

We measure the spin polarization by periodically inverting the nuclei
of choice using adiabatic rapid passages
\cite{PoggioAPL:2007,Degen:2007}.  We operate in a fixed magnetic
field ${\bf B}_0={\bf B}_{ext}+{\bf B}_{tip}$ ($|{\bf B}_0| = 2.72$
T), where ${\bf B}_{tip}$ is the field produced by the magnetic tip.
We periodically sweep the frequency $\nu_I$ of a transverse rf
magnetic field ${\bf B}_{1I}$ through the Larmor resonance condition,
$\nu_{I} = \frac{\gamma_I}{2 \pi}B_0$ so as to induce adiabatic
inversions of the nuclear spin polarization.  In the presence of the
magnetic tip, periodic inversions of the spin polarization generate an
alternating force that drives the mechanical resonance of the
cantilever.  The amplitude of cantilever oscillation, which we measure
using optical interferometry, is then proportional to the $I$-spin
polarization.  As shown in Fig.~\ref{fig1}, we can measure either the
$^1$H or the $^{13}$C statistical polarization by adjusting the rf
center frequency of the adiabatic passages.

\begin{figure}\includegraphics{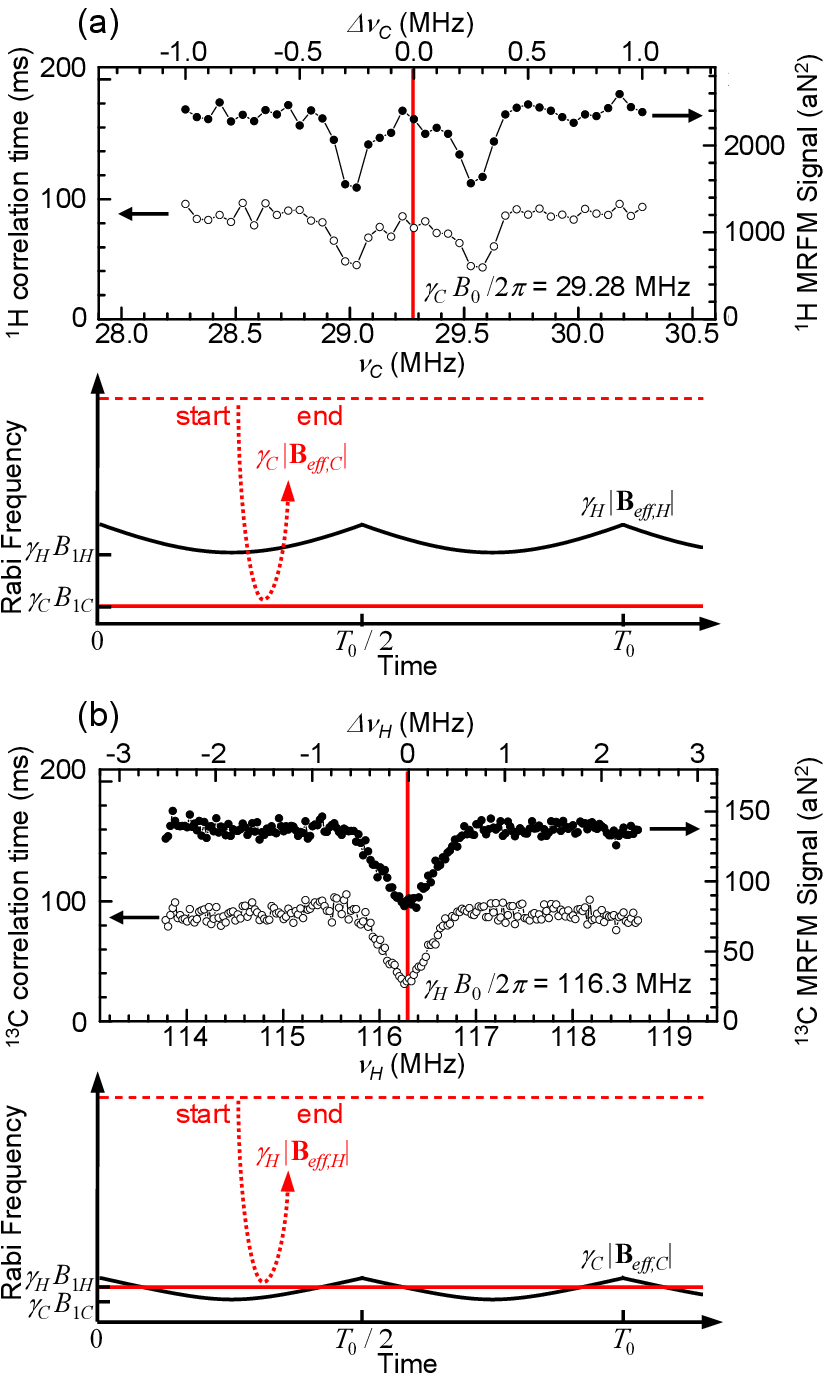}
\caption{\label{fig3}
  CP resonances detected by MRFM in statistical polarizations plotted
  as a function of the $S$-spin resonance offset.  Below each plot,
  the corresponding trajectories of the effective field Rabi
  frequencies are shown.  The effect of scanning $\Delta \nu_S$ is to
  vertically shift $\gamma_S |\BeffS|$ as shown by the dotted arrow.
  (a) The ``observed'' spin $I =$ $^1$H and the ``unobserved'' spin $S
  =$ $^{13}$C with $\nu_{FM} = 300$ kHz.  (b) The ``observed'' spin $I
  =$ $^{13}$C and the ``unobserved'' spin $S =$ $^1$H with $\nu_{FM} =
  125$ kHz.  }\end{figure}

In order to observe polarization transfer, we perform adiabatic
passages on one isotope --- the ``observed'' or $I$-spin isotope.
This measurement produces a signal that is proportional to the
fluctuating $I$-spin polarization.  Simultaneously, we address the
other isotope --- the ``unobserved'' or $S$-spin isotope --- with cw
radiation at or near its Larmor frequency.  This continuously applied
resonant $\BS$ is constant in the rotating frame and remains locked to
a statistical polarization of $S$ spins with a correlation time
$\tmS$.  The basic scheme is shown in Fig.~\ref{fig2}.

\begin{figure}\includegraphics{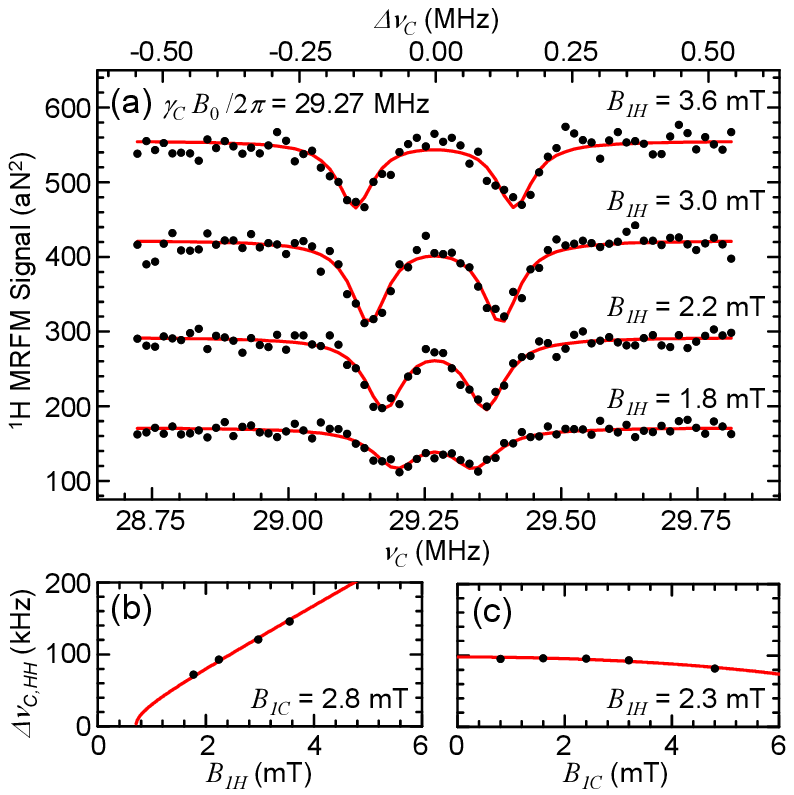}
\caption{\label{fig4}
  (a) MRFM data (points) and fits (lines) as a function of the
  $S$-spin resonance offset for $I =$ $^1$H and $S =$ $^{13}$C for
  different $B_{1H}$, with $B_{1C}$ = 2.8 mT\@.  (b) $\Delta
  \nu_{C,HH}$ plotted as points for different $B_{1H}$ taken from fits
  in (a).  (c) $\Delta \nu_{C,HH}$ for different $B_{1C}$ from fits to
  a separate data set (not shown).  Lines in (b) and (c) correspond to
  (\ref{eq2}) with no free parameters.  }\end{figure}

When the HH condition is fulfilled, polarization transfer occurs
between the statistically polarized ``observed'' and ``unobserved''
spin ensembles.  Measurements are shown in Fig.~\ref{fig3} for both
$^1$H and $^{13}$C in the role of the ``observed'' isotope.  In both
experiments we increment the resonance offset $\Delta \nu_S$ of the
$S$ spins while detecting the $I$ spins with frequency sweeps centered
on $\Delta \nu_I = 0$.  We record both the signal variance and the
correlation time of the $I$-spin fluctuations.  Dips in the
correlation time appear at those frequencies where $\Delta \nu_S$
satisfies the HH condition.  Since the signal is recorded in a narrow
band ($\sim 17$ Hz) around the cantilever resonance, a reduction in
the correlation time due to CP also gives rise to a reduction in the
observed force signal.

Fig.~\ref{fig3} shows two distinct CP regimes.  In the first regime,
shown in Fig.~\ref{fig3}(a), we observe the $^1$H spins, using
$\gamma_H B_{1H} = 280$ kHz and $\gamma_C B_{1C} = 29$ kHz, so that
$\gamma_I \BI > \gamma_S \BS$.  Again, as we increment $\Delta \nu_C$,
$\gamma_C |\BeffC|$ comes to a minimum at resonance according to
$\gamma_C |\BeffC| = \sqrt{(\gamma_C B_{1C})^2 + (2 \pi \Delta
  \nu_C)^2}$.  For two bands of $\Delta \nu_C$ symmetric about zero,
$\gamma_C |\BeffC|$ intersects the trajectory of $\gamma_H |\BeffH|$
thereby fulfilling the HH condition \cite{Eberhardt:2007}.  The
resulting CP produces the double-dip structure shown in
Fig.~\ref{fig3}(a).  The most efficient CP and therefore the most
significant reduction in the $^1$H correlation time (\ie the deepest
part of the dips) occurs for a HH match at the vertex of the $\gamma_H
| \BeffH |$ hyperbola shown in Fig.~\ref{fig3}(a).  At these
intersections the slopes of $\gamma_C | \BeffC |$ and $\gamma_H |
\BeffH |$ match, producing the longest possible HH contact.  In
addition, CP is most efficient when $\Delta \nu_C$ and $\Delta \nu_H$
are smallest resulting in small angles of the effective field.

In the second regime, shown in Fig.~\ref{fig3}(b), we observe the
$^{13}$C spins using $\gamma_H B_{1H} = 120$ kHz and $\gamma_C B_{1C}
= 62$ kHz, so that $\gamma_I \BI < \gamma_S \BS$.  As we increment
$\Delta \nu_H$, $\gamma_H |\BeffH|$ comes to a minimum at resonance
according to $\gamma_H |\BeffH| = \sqrt{(\gamma_H B_{1H})^2 + (2 \pi
  \Delta \nu_H)^2}$.  In this case, there is one band of $\Delta
\nu_H$ symmetric about zero for which $\gamma_H |\BeffH |$ intersects
the trajectory of $\gamma_C | \BeffC |$. The resulting CP produces the
single-dip structure of Fig.~\ref{fig3}(b).  For the same reasons
which apply in the first regime, the most efficient CP occurs for
$\Delta \nu_H = 0$ where the HH match is closest to the vertex of the
$\gamma_C | \BeffC |$ hyperbola.  The same single- and double-dip
behavior shown in Fig.~\ref{fig3} appears in Fig. 2 of the original
paper on CP by Hartmann and Hahn \cite{Hartmann:1962}.

We have measured a series of spectra of the double-dip type for
different $\gamma_H B_{1H}$ and $\gamma_C B_{1C}$.  Typical results
are shown in Fig.~\ref{fig4}(a).  Using simple Lorentzian fits to
determine the position of the dips, we extract the splitting for all
spectra.  From the condition $\gamma_{I} \left| \BeffI \right| =
\gamma_{S} \left | \BeffS \right|$ and as pointed out by Hartmann and
Hahn (Ref.  \cite{Hartmann:1962}, Eq.  (75)), the dips in
Fig.~\ref{fig4}(a) should appear at
\begin{equation}
\label{eq2}
\Delta \nu_C = \pm \Delta \nu_{C, HH} \equiv \pm \sqrt{\frac{\gamma_H^2}{4 \pi^2} B_{1H}^2 -
\frac{\gamma_C^2}{4 \pi^2} B_{1C}^2},
\end{equation}
so that the splitting is $2\Delta \nu_{C,HH}$.  The experimental data
in Fig.~\ref{fig4}(b) and (c) agree within the error with the
theoretical curve representing (\ref{eq2}) without any adjustable
parameters.  Note that the magnitudes of $B_{1H}$ and $B_{1C}$ are
calibrated by independent nutation experiments on each species
\cite{PoggioAPL:2007}.  Given this agreement, we confirm that our
double-resonance features result from the HH effect.

While in our experiments it was possible to observe the $^{13}$C
signal directly, double-resonance detection is particularly useful
when one isotope with spin $S$ has a weak resonance signal, either
because of a small $\gamma_S$ or a low abundance.  The presence of the
$S$ spins can then be detected via the stronger signal of the $I$
spins.  The ability to perform CP in statistical ensembles provides
the possibility of new contrast mechanisms for nanoscale MRI
applications.  For example, organic material with many proximate
$^{13}$C and $^1$H atoms could be distinguished from interstitial
water molecules, which possess only $^1$H.  Statistical double
resonance could be combined with advanced spectroscopy techniques for
chemical characterization of materials at the nanometer-scale.
Ultimately, such techniques could be applied to discern individual
protein components in complex nanometer-scale biological structures.

\begin{acknowledgments}
We acknowledge support from the NSF-funded Center for Probing the
Nanoscale (CPN) at Stanford University.
\end{acknowledgments}


\begin{thebibliography}{23}
\bibitem{NoteTauCP} This estimate results from the rms average of C-H
  dipolar half-splittings in a CH$_2$ group taken over a powder
  distribution of field orientations.  We use a C-H bond length of 1.1
  $\rm \AA$ and a H-C-H angle of 109.5$^{\circ}$ for the calculation.
\bibitem{footnote} When the total polarization $\DNI + \DNS = 0$, the
  probability distribution for a given value of $\DNI$ is given by the
  product of two binomial distributions, which can be approximated by
  Gaussians with variances $\NI$ and $\NS$ respectively.  The
  resulting distribution is another Gaussian with variance $\NI \NS /
  (\NI + \NS)$.
\end{thebibliography}
\end{document}